\documentclass[12pt,a4paper]{article}
\usepackage{blkarray}
\usepackage{amsfonts}
\usepackage{amsmath}
\usepackage{geometry}
\usepackage{graphicx}
\usepackage{amssymb}
\usepackage{abstract}
\usepackage{setspace}
\usepackage{geometry}
\providecommand{\U}[1]{\protect\rule{.1in}{.1in}}

\def\fr{\frac}
\def\be{\begin{equation}}
\def\ee{\end{equation}}
\def\ba{\begin{eqnarray}}
\def\ea{\end{eqnarray}}
\def\pa{\partial}
\def\ra{\rightarrow}

\def\g{\gamma}

\def\pt{\phantom{a}}

\begin{document}

\title{3 fermionic families naturally arise from hyperionic formalism}

\author{Diego Marin \thanks{dmarin.math@gmail.com}}

\maketitle

\begin{abstract}\begin{spacing}{1.2}
\noindent We'll show as a quantum theory based on hyperions (said also \lq complex quaternions') gives reason straightforwardly for the existence of three fermionic families, dark matter and superfields which don't require never seen particles.
\end{spacing}
\end{abstract}

\newpage
\tableofcontents
\newpage

\section{Hyperions}

We give briefly a definition which we'll use repeatedly. \textbf{Hyperions}, sometimes called \lq complex quaternions' are hyper-complex numbers with seven imaginary unities $i,j,k,I,iI,jI,kI$ and commutation rules:

$$ij = -ji = k \qquad jk = -kj = i \qquad ki = -ik = j \qquad i^2 = j^2 = k^2 = I^2 = -1$$
$$iI = Ii \qquad jI = Ij \qquad kI = Ik \qquad (iI)^2 = (jI)^2 = (kI)^2 = 1$$

\noindent The algebra so defined is associative.

\section{Introduction}

It is well known that in a grand-unified theory with $SU(6)$ as gauge group, all fermionic fields can be arranged in a skew-symmetric matrix $\chi$ which transforms in the skew-symmetric representation. Every entry in the matrix will be a Dirac spinor:

\be \chi \ra e^{iT} \chi e^{iT*} \qquad \text{with} \quad T \in su(6)\label{tras1}\ee

\noindent This formalism doesn't give any explanation about the experimentally verified existence of three fermionic families which are identical except for masses. Moreover, the transformation law (\ref{tras1}) makes very hard an unification with bosonic fields $F$ which transform in the adjoint representation:

\be F \ra e^{iT} F e^{-iT} \qquad \text{with} \quad T \in su(6)\label{tras2}\ee

\noindent In what follows we demonstrate that a generic hyperionic field $\psi$, which transforms in the adjoint representation, takes into account for three fermionic families which transform in the skew symmetric reprentation.

\section{From \lq adjoint' to \lq skew-symmetric'}

Without loss of generality, we can write a hyperionic field as

$$\psi = \psi_0 + \psi_1 \underline{k} + \psi_2 \underline{i} + \psi_3 \underline{j}$$

\noindent with

\ba \psi_0 &=& \psi_0^1 + I\psi_0^2 \nonumber \\
    \psi_1 &=& \phi_1^1 - \underline{i}\xi_1^1 + I(\phi_1^2 -\underline{i}\xi_1^2) \nonumber \\
    \psi_2 &=& \phi_2^1 - \underline{j}\xi_2^1 + I(\phi_2^2 -\underline{j}\xi_2^2) \nonumber \\
    \psi_3 &=& \phi_3^1 - \underline{k}\xi_3^1 + I(\phi_3^2 -\underline{k}\xi_3^2) \nonumber \\
    && \!\!\!\!\!\!\!\!\!\!\!\!\!\!\!\!\!\!\!\!
    \phi_m^n, \xi_m^n \in \mathbb{R} \qquad m=1,2,3; \pt n=1,2 \nonumber \ea

\noindent We have three ways to construct $SU(6)$ transformations

$$ [SU(6)]^{(1)} \approx e^{\underline{i}T} \qquad [SU(6)]^{(2)} \approx e^{\underline{j}T'} \qquad [SU(6)]^{(3)} \approx e^{\underline{k}T''} $$

\noindent where $T$ contains only $\underline{i}$ as imaginary unity, $T'$ contains only $\underline{j}$ and $T''$ contains only $\underline{k}$. Consider now a hyperionic field $\psi$ which transforms in the adjoint representation for any class of $SU(6)$:

\ba && [SU(6)]^{(1)} \approx e^{\underline{i}T} \nonumber \\
    && \psi \ra \underbrace{e^{\underline{i}T} \psi e^{-\underline{i}T}}_{Adjoint\,\,rep} \Rightarrow \psi_1 \underline{k} \ra e^{\underline{i}T}\psi_1 \underline{k} e^{-\underline{i}T} = e^{\underline{i}T} \psi_1 e^{\underline{i}T*} \underline{k} \nonumber \\
    && \pt\qquad\qquad\qquad \Rightarrow \psi_1 \ra\underbrace{e^{\underline{i}T} \psi_1 e^{\underline{i}T*}}_{Skew\,\,sym\,\,rep}\qquad\qquad \left( \begin{array}{l} \text{$T$ and $\psi_1$ contain}\\
                        \text{only $\underline{i}$ as imagina-}\\
                        \text{ry unit}\end{array}\right)\nonumber \ea

\ba && [SU(6)]^{(2)} \approx e^{\underline{j}T'} \nonumber \\
    && \psi \ra \underbrace{e^{\underline{j}T'} \psi e^{-\underline{j}T'}}_{Adjoint\,\,rep} \Rightarrow \psi_2 \underline{i} \ra e^{\underline{j}T'}\psi_2 \underline{i} e^{-\underline{j}T'} = e^{\underline{j}T'} \psi_2 e^{\underline{j}T'*} \underline{i} \nonumber \\
    && \pt\qquad\qquad\qquad \Rightarrow \psi_2 \ra\underbrace{e^{\underline{j}T'} \psi_2 e^{\underline{j}T'*}}_{Skew\,\,sym\,\,rep}\qquad\qquad \left( \begin{array}{l}
    \text{$T'$ and $\psi_2$ contain}\\ \text{only $\underline{j}$ as imagina-}\\
    \text{ry unit}\end{array}\right)\nonumber \ea

\ba && [SU(6)]^{(3)} \approx e^{\underline{k}T''} \nonumber \\
    && \psi \ra \underbrace{e^{\underline{k}T''} \psi e^{-\underline{k}T''}}_{Adjoint\,\,rep} \Rightarrow \psi_3 \underline{j} \ra e^{\underline{k}T''}\psi_3 \underline{j} e^{-\underline{k}T''} = e^{\underline{k}T''} \psi_3 e^{\underline{k}T''*} \underline{j} \nonumber \\
    && \pt\qquad\qquad\qquad \Rightarrow \psi_3 \ra\underbrace{e^{\underline{k}T''} \psi_3 e^{\underline{k}T''*}}_{Skew\,\,sym\,\,rep}\qquad\qquad \left( \begin{array}{l} \text{$T''$ and $\psi_3$ contain}\\ \text{only $\underline{k}$ as imagina-}\\ \text{ry unit}\end{array}\right)\nonumber \ea

\noindent In all the cases:

\ba \psi_0 &\overset{(1)}{\ra}& e^{\underline{i}T}\psi_0 e^{-\underline{i}T} = \psi_0' \nonumber \\
    \psi_0 &\overset{(2)}{\ra}& e^{\underline{j}T'}\psi_0 e^{-\underline{j}T'} = \psi_0'' \nonumber \\
    \psi_0 &\overset{(3)}{\ra}& e^{\underline{k}T''}\psi_0 e^{-\underline{k}T''} = \psi_0''' \nonumber \ea

\noindent where $\psi_0',\psi_0'',\psi_0'''$ contain only $I$ as imaginary unit. We can say that transformations in $SU(6)$ maintain $\psi_0$ in its family, i.e. it doesn't interact with ordinary matter via $SU(6)$ gauge fields. This means that $\psi_0$ is a good candidate for dark matter. Conversely, if we apply a transformation in $[SU(6)]^{(n)}$ to $\psi_m$, with $m\neq n$, we obtain a field $\psi_q$ in the remaining family $q \neq m,n$.

\section{Gamma representation}

What remains to be demonstrated is the equality between a hyperionic field and three families of spinor (Dirac) fields. Let's start by considering the correspondence between hyperionic imaginary unities and gamma matrices. Explicitly:

$$\underline{i} = \g_2\g_1 \qquad \underline{j} = \g_1\g_3 \qquad \underline{k} = \g_3\g_2 \qquad I = \g_0\g_1\g_2\g_3$$
$$\underline{i}I=I\underline{i}=\g_0\g_3 \qquad \underline{j}I = I\underline{j} = \g_0\g_2 \qquad \underline{k}I = I\underline{k} = \g_0\g_1$$

\noindent We take gamma matrices in Dirac representation:

$$\g_0 = \left(\begin{array}{cccc} -1 &  0 & 0 & 0 \\
                                    0 & -1 & 0 & 0 \\
                                    0 &  0 & 1 & 0 \\
                                    0 &  0 & 0 & 1 \end{array}\right) \qquad
\g_1 = \left(\begin{array}{cccc} 0 &  0 & 0 & 1 \\
                                 0 &  0 & 1 & 0 \\
                                 0 & -1 & 0 & 0 \\
                                -1 &  0 & 0 & 0 \end{array}\right) $$
$$\g_2 = \left(\begin{array}{cccc} 0 & 0 & 0 & -i \\
                                   0 & 0 & i &  0 \\
                                   0 & i & 0 &  0 \\
                                  -i & 0 & 0 &  0 \end{array}\right) \qquad
\g_3 = \left(\begin{array}{cccc} 0 & 0 & 1 &  0 \\
                                 0 & 0 & 0 & -1 \\
                                -1 & 0 & 0 &  0 \\
                                 0 & 1 & 0 &  0 \end{array}\right) $$

\noindent This gives:

$$\underline{i} = \left(\begin{array}{cccc} i &  0 & 0 &  0 \\
                                            0 & -i & 0 &  0 \\
                                            0 &  0 & i &  0 \\
                                            0 &  0 & 0 & -i \end{array}\right) \qquad
\underline{j} = \left(\begin{array}{cccc} 0 & 1 &  0 & 0 \\
                                         -1 & 0 &  0 & 0 \\
                                          0 & 0 &  0 & 1 \\
                                          0 & 0 & -1 & 0 \end{array}\right) $$
$$\underline{k} = \left(\begin{array}{cccc} 0 & i & 0 & 0 \\
                                            i & 0 & 0 & 0 \\
                                            0 & 0 & 0 & i \\
                                            0 & 0 & i & 0 \end{array}\right) \qquad
\underline{k}I = \left(\begin{array}{cccc} 0 &  0 &  0 & -1 \\
                                           0 &  0 & -1 &  0 \\
                                           0 & -1 &  0 &  0 \\
                                          -1 &  0 &  0 &  0 \end{array}\right) $$
$$I = -\underline{k}I\underline{k} = \left(\begin{array}{cccc} 0 & 0 & i & 0 \\
                                                               0 & 0 & 0 & i \\
                                                               i & 0 & 0 & 0 \\
                                                               0 & i & 0 & 0 \end{array}\right) \qquad
\underline{i}I = \left(\begin{array}{cccc} 0 & 0 & -1 & 0 \\
                                           0 & 0 &  0 & 1 \\
                                          -1 & 0 &  0 & 0 \\
                                           0 & 1 &  0 &  0 \end{array}\right) $$
$$\underline{j}I = \left(\begin{array}{cccc} 0 &  0 &  0 & i \\
                                             0 &  0 & -i & 0 \\
                                             0 &  i &  0 & 0 \\
                                            -i &  0 &  0 & 0 \end{array}\right) $$

\noindent By expanding $\psi$:

\ba\psi = \psi_0^1 +I\psi_0^2 +\underline{k}\phi_1^1 +\underline{j}\xi_1^1 +\underline{k}I\phi_1^2 +\underline{j}I\xi_1^2 &&\nonumber \\
+ \underline{i}\phi_2^1 +\underline{k}\xi_2^1 +\underline{i}I\phi_2^2 +\underline{k}I\xi_2^2 &&\nonumber \\
+ \underline{j}\phi_3^1 + \underline{i}\xi_3^1 +\underline{j}I \phi_3^2 + \underline{i}I \xi_3^2 &&\nonumber \ea

$$ \psi = \left( \begin{array}[c]{c|c|}
\psi_0^1 + i(\phi^1_2+\xi_3^1)&(\xi_1^1+\phi_3^1) +i(\phi^1_1 +\xi^1_2) \\
i(\phi^1_1 +\xi^1_2)-(\xi_1^1+\phi_3^1)&\psi_0^1-i(\phi^1_2+\xi_3^1) \\
i\psi_0^2-(\phi^2_2+\xi_3^2)&-(\phi^2_1 +\xi^2_2)+i(\xi_1^2+\phi_3^2) \\
-(\phi^2_1 +\xi^2_2)-i(\xi_1^2+\phi_3^2)& (\phi^2_2+\xi_3^2)+i\psi_0^2 \\
\end{array}\right. $$
$$ \left. \begin{array}[c]{|c|c}
i\psi_0^2-(\phi^2_2+\xi_3^2) & -(\phi^2_1 +\xi^2_2)+i(\xi_1^2+\phi_3^2)   \\
-(\phi^2_1 +\xi^2_2)-i(\xi_1^2+\phi_3^2) &(\phi^2_2+\xi_3^2)+i\psi_0^2    \\
i(\phi^1_2+\xi_3^1)+\psi_0^1 &(\xi_1^1+\phi_3^1)+i(\phi^1_1 +\xi^1_2)     \\
-(\xi_1^1+\phi_3^1)+i(\phi^1_1 +\xi^1_2) &-i(\phi^1_2+\xi_3^1)+\psi_0^1   \\
\end{array}\right) $$

\noindent The structure is then

$$\psi = \left(\begin{array}{c|c|c|c} a & -b^* & c &  d^* \\
                                      b &  a^* & d & -c^* \\
                                      c &  d^* & a & -b^* \\
                                      d & -c^* & b &  a^* \end{array}\right) \qquad
\psi^\dag = \left(\begin{array}{c|c|c|c} a^* &  b^* &  c^* &  d^* \\
                                        -b   &  a   &  d   & -c   \\
                                         c^* &  d^* &  a^* &  b^* \\
                                         d   & -c   & -b   &  a   \end{array}\right) $$

\noindent Consider now the Dirac derivative operator (noting that $\g_0\g^0 = -(\g_0)^2 = -\mathbf{1}$):

$$M = \g_0 \g^\mu \pa_\mu = \left(\begin{array}{c|c|c|c} -\pa_0 &  0 &  -\pa_3 &  -\pa_1+i\pa_2 \\
                                        0   &  -\pa_0   &  -\pa_1-i\pa_2   & \pa_3   \\
                                         -\pa_3 &  -\pa_1+i\pa_2 &  \pa_0 &  0 \\
                                         -\pa_1-i\pa_2   & \pa_3   & 0   &  \pa_0   \end{array}\right) $$

\noindent We see that $M$ uses only four unities ($1 = \g_0\g_0$, $\underline{k}I = \g_0\g_1$, $\underline{j}I = \g_0\g_2$, $\underline{i}I = \g_0\g_3$) omitting $I,\underline{i},\underline{j},\underline{k}$. The complete extension is obtained by introducing complex dimensions

$$x^\mu = Re(x^\mu)+I\,Im(x^\mu) = Re(x^\mu)+\g_0\g_1\g_2\g_3\,Im(x^\mu)$$

\noindent and complex derivatives

$$\fr \pa{\pa x^\mu} = \fr \pa{\pa Re(x^\mu)}-I\,\fr\pa{\pa Im(x^\mu)} = \fr \pa{\pa Re(x^\mu)}-\g_0\g_1\g_2\g_3\,\fr\pa{\pa Im(x^\mu)}$$

\noindent However we proceed with the simpler case, leaving the complete one to future works.

Our target now is to introduce a new lagrangian $\mathfrak{L}= Re(\psi^\dag M \psi)$ (or $\mathfrak{L}=Tr(\psi^\dag M \psi)$ in the Gamma-representation), demonstrating its equivalence with ordinary Dirac lagrangian $\mathfrak{L} = \chi^\dag M \chi +c.c.$ for some ordinary Dirac spinor $\chi$.

$$ M\psi = \left( \begin{array}[c]{c|c|}
-\pa_0 a -\pa_3 c -\pa_1 d +i\pa_2 d & \pa_0 b^* -\pa_3 d^* +\pa_1 c^* -i\pa_2 c^* \\
-\pa_0 b -\pa_1 c -i\pa_2 c +\pa_3 d & -\pa_0 a^* -\pa_1 d^* -i\pa_2 d^* -\pa_3 c^* \\
-\pa_3 a -\pa_1 b +i\pa_2 b +\pa_0 c & \pa_3 b^* -\pa_1 a^* +i\pa_2 a^* +\pa_0 d^* \\
-\pa_1 a -i\pa_2 a +\pa_3 b +\pa_0 d & \pa_1 b^* + i\pa_2 b^* +\pa_3 a^* -\pa_0 c^* \\
\end{array}\right. $$
$$ \left. \begin{array}[c]{|c|c}
-\pa_0 c -\pa_3 a -\pa_1 b +i\pa_2 b & -\pa_0 d^* + \pa_3 b^* -\pa_1 a^* +i\pa_2 a^*  \\
-\pa_0 d -\pa_1 a -i\pa_2 a +\pa_3 b & \pa_0 c^* +\pa_1 b^* + i\pa_2 b^* +\pa_3 a^* \\
-\pa_3 c -\pa_1 d +i\pa_2 d +\pa_0 a & -\pa_3 d^* +\pa_1 c^* -i\pa_2 c^* -\pa_0 b^* \\
-\pa_1 c -i\pa_2 c +\pa_3 d +\pa_0 b & -\pa_1 d^* -i\pa_2 d^* -\pa_3 c^* +\pa_0 a^* \\
\end{array}\right) $$

\noindent Consider now the identity $Tr(\psi^\dag M\psi) = (\psi^\dag M\psi)^{00} + (\psi^\dag M\psi)^{11}+ (\psi^\dag M\psi)^{22}+ (\psi^\dag M\psi)^{33}$ and proceed one term at a time:

\ba (\psi^\dag M \psi)^{00} &=& -a^* \pa_0 a - a^*\pa_3 c - a^*\pa_1 d + ia^*\pa_2 d \nonumber \\
                            && -b^*\pa_0 b -b^*\pa_1 c -ib^*\pa_2 c +b^*\pa_3 d \nonumber \\
                            && -c^*\pa_3 a -c^*\pa_1 b +ic^*\pa_2 b +c^*\pa_0 c \nonumber \\
                            && -d^*\pa_1 a -id^*\pa_2 a +d^*\pa_3 b + d^*\pa_0 d \nonumber \ea

\ba (\psi^\dag M \psi)^{11} &=& -b\pa_0 b^* + b\pa_3 d^* - b\pa_1 c^* + i\pa_2 c^* \nonumber \\
                            && -a\pa_0 a^* -a\pa_1 d^* -ia\pa_2 d^* -a\pa_3 c^* \nonumber \\
                            && +d\pa_3 b^* -d\pa_1 a^* + id\pa_2 a^* +d\pa_0 d^* \nonumber \\
                            && -c\pa_1 b^* -ic\pa_2 b^* -c\pa_3 a + c\pa_0 c^* \nonumber \ea

\ba (\psi^\dag M \psi)^{22} &=& -c^*\pa_0 c -c^*\pa_3 a -c^*\pa_1 b +ic^*\pa_2 b\nonumber \\
                            && -d^*\pa_0 d - d^* \pa_1 a -id^*\pa_2 a +d^*\pa_3 b \nonumber \\
                            && -a^*\pa_3 c -a^*\pa_1 d +ia^*\pa_2 d +a^*\pa_0 a \nonumber \\
                            && -b^*\pa_1 c -ib^*\pa_2 c +b^*\pa_3 d +b^*\pa_0 b \nonumber \ea

\ba (\psi^\dag M \psi)^{33} &=& -d\pa_0 d^* + d\pa_3 b^* - d\pa_1 a^* + id\pa_2 a^* \nonumber \\
                            && -c\pa_0 c^* - c\pa_1 b^* - ic \pa_2 b^* -c\pa_3 a^* \nonumber \\
                            && +b\pa_3 d^* -b\pa_1 c^* + ib\pa_2 c^* + b\pa_0 b^* \nonumber \\
                            && -a\pa_1 d^* -ia\pa_2 d^* - a\pa_3 c^* + a\pa_0 a^* \nonumber \ea

\noindent Hence:

\ba Tr(\psi^\dag M \psi) &=& -2a^*\pa_3 c - 2a^*\pa_1 d +2ia^*\pa_2 d -2b^*\pa_1 c \nonumber \\
                         && -2ib^*\pa_2 c +2b^*\pa_3 d -2c^*\pa_3 a -2c^*\pa_1 b \nonumber \\
                         && +2ic^*\pa_2 b -2d^*\pa_1 a -2id^*\pa_2 a + 2d^*\pa_3 b \nonumber \\
                         && +2b\pa_3 d^* -2b\pa_1 c^* +2ib\pa_2 c^* -2a\pa_1 d^* \nonumber \\
                         && -2ia\pa_2 d^* -2a\pa_3 c^* + 2d\pa_3 b^* -2d\pa_1 a^* \nonumber \\
                         && +2id\pa_2 a^* -2c\pa_1 b^* - 2ic\pa_2 b^* - 2c\pa_3 a^* \nonumber \ea

\noindent Let's repeat calculation for Dirac lagrangian:

$$ M\chi = \left( \begin{array}[c]{c}
-\pa_0 \chi_0 -\pa_3 \chi_2 -\pa_1 \chi_3 +i\pa_2\chi_3 \\
-\pa_0 \chi_1 -\pa_1 \chi_2 -i\pa_2 \chi_2 +\pa_3\chi_3 \\
-\pa_3 \chi_0 -\pa_1\chi_1 +i\pa_2\chi_1 +\pa_0\chi_2 \\
-\pa_1 \chi_0 -i\pa_2\chi_0 +\pa_3\chi_1 +\pa_0\chi_3
\end{array}\right) $$

\ba \chi^\dag M\chi &=& -\chi_0^*\pa_0 \chi_0 - \chi_0^*\pa_3 \chi_2 -\chi_0^*\pa_1 \chi_3 +i\chi_0^*\pa_2\chi_3 \nonumber \\
&& -\chi_1^* \pa_0 \chi_1 - \chi_1^*\pa_1\chi_2 -i\chi_1^*\pa_2\chi_2 +\chi_1^*\pa_3\chi_3 \nonumber \\
&& -\chi_2^* \pa_3\chi_0 - \chi_2^*\pa_1\chi_1 +i\chi_2^*\pa_2\chi_1 + \chi_2^*\pa_0\chi_2 \nonumber \\
&& -\chi_3^* \pa_1 \chi_0 -i\chi_3^*\pa_2\chi_0 +\chi_3^*\pa_3\chi_1 +\chi_3^*\pa_0 \chi_3 \nonumber \ea

\ba \chi^\dag M\chi +c.c. &=& - \chi_0^*\pa_3 \chi_2 -\chi_0^*\pa_1 \chi_3 +i\chi_0^*\pa_2\chi_3 \nonumber \\
&& - \chi_1^*\pa_1\chi_2 -i\chi_1^*\pa_2\chi_2 +\chi_1^*\pa_3\chi_3 \nonumber \\
&& -\chi_2^* \pa_3\chi_0 - \chi_2^*\pa_1\chi_1 +i\chi_2^*\pa_2\chi_1 \nonumber \\
&& -\chi_3^* \pa_1 \chi_0 -i\chi_3^*\pa_2\chi_0 +\chi_3^*\pa_3\chi_1 \nonumber \\
&& - \chi_0\pa_3 \chi_2^* -\chi_0\pa_1 \chi_3^* -i\chi_0\pa_2\chi_3^* \nonumber \\
&& - \chi_1\pa_1\chi_2^* +i\chi_1\pa_2\chi_2^* +\chi_1\pa_3\chi_3^* \nonumber \\
&& -\chi_2 \pa_3\chi_0^* - \chi_2\pa_1\chi_1^* -i\chi_2\pa_2\chi_1^* \nonumber \\
&& -\chi_3\pa_1 \chi_0^* +i\chi_3\pa_2\chi_0^* +\chi_3\pa_3\chi_1^* \nonumber \ea

\noindent A quick comparison leads to the following result:

$$ \chi = {\sqrt 2}\left( \begin{array}[c]{c}
a \\ b \\ c \\ d
\end{array}\right)= {\sqrt 2}\left( \begin{array}[c]{c}
\psi_0^1+i(\phi_2^1+\xi_3^1) \\
-(\xi_1^1+\phi_3^1)+i(\phi_1^1+\xi_2^1) \\
-(\phi_2^2+\xi_3^2)+i\psi_0^2 \\
-(\phi_1^2+\xi_2^2)-i(\xi_1^2+\phi_3^2)
\end{array}\right) $$

\noindent So we have \\
$\pt$\\
FAMILY-1 $\pt\qquad\qquad\qquad\qquad\qquad\qquad\pt$ FAMILY-2

$$ \chi^{(1)} = {\sqrt 2}\left( \begin{array}[c]{c}
0 \\
-\xi_1^1 +i\phi_1^1 \\
0 \\
-\phi_1^2 -i\xi_1^2
\end{array}\right) \qquad\qquad\qquad\qquad
\chi^{(2)} = {\sqrt 2}\left( \begin{array}[c]{c}
i\phi_2^1 \\
i\xi_2^1 \\
-\phi_2^2 \\
-\xi_2^2
\end{array}\right)$$

\noindent FAMILY-3 $\pt\qquad\qquad\qquad\qquad\qquad\qquad\pt$ DARK-MATTER

$$ \chi^{(3)} = {\sqrt 2}\left( \begin{array}[c]{c}
i\xi_3^1 \\
-\phi_3^1 \\
-\xi_3^2 \\
-i\phi_3^2
\end{array}\right) \qquad\qquad\qquad\qquad
\chi^{(DM)} = {\sqrt 2}\left( \begin{array}[c]{c}
\psi_0^1 \\
0 \\
i\psi_0^2 \\
0
\end{array}\right)$$

\section{Superfield}

Let consider gauge fields $A_\mu$ for gauge transformations in $U(6,\mathbf{H})$, where a matrix $U$ belongs to $U(6,\mathbf{H})$ if and only if

$$U\overline{U}^\dag = \overline{U}^\dag U = 1$$

\noindent where $\dag$ represents ordinary hermitian conjugation (transposition plus complex conjugation), while $-$ is an extra conjugation restricted to imaginary unit $I$\footnote{Explicitly $i^* = -i$, $j^* =-j$, $k^* = -k$, $I^* = -I$, $(iI)^*=iI$, $(jI)^* = jI$, $(kI)^* = kI$, $\overline{i} = i$, $\overline{j} =j$, $\overline{k} = k$, $\overline{I} = -I$, $\overline{(iI)}=-iI$, $\overline{(jI)} = -jI$, $\overline{(kI)} = -kI$.}. In the Cartan classification this group corresponds to $Sp(12,\mathbf{C})$.

Fields $A_\mu$ are real linear combinations of elements in the corresponding Lie algebra $u(6,\mathbf{H})$, i.e. $\overline{A}^\dag_\mu = -A_\mu$. The same algebra is filled exactly by all the already known fermionic fields, pigeonholed in three families inside $\psi$. Note that

$$Tr\,u(6,\mathbf{H}) = so(1,3) \Longleftrightarrow Det\,U(6,\mathbf{H}) = SO(1,3)$$

\noindent This means that $U(6,\mathbf{H})$ includes $SO(1,3)$, i.e. the gauge group of gravity! The six generators are obviously

$$i(\g_2\g_1), j(\g_1\g_3), k(\g_3\g_2), iI(\g_0\g_3), jI(\g_0\g_2), kI(\g_0\g_1)$$

\noindent Moreover, is precisely $SO(1,3)$ (hence gravity) which regulates transitions between homologous fields in different families.

Being both $A_\mu$ and $\psi$ in the adjoint representation, we can join them in a unique superfield:

$$\Psi = e^\mu A_\mu + \theta \psi$$

\noindent $e^\mu$ is an hyperionic tetrad for the ordinary space. The internal dimensions are in one to one correspondence with the unities $1,iI,jI,kI$ (or with all the unities $1,i,j,k,I,iI,jI,kI$ if we consider complex dimensions). Thus the internal index disappear. $\theta$ is the analogous of $e^\mu$ for one grassmannian dimension.

\section{Conclusion}

By concluding, we can say that hyperionic formalism not only justifies the existence of three fermionic failies, but also it permits an easy unification with gauge fields and dark matter.


\newpage







\begin{thebibliography}{}

\bibitem {Arrangement} Marin, D.: Arrangement Field Theory: beyond Strings and Loop Gravity. LAP LAMBERT Academic Publishing (August 31, 2012)

\bibitem {Quat1} De Leo, S., Ducati, G.: Quaternionic differential operators. Journal of Mathematical Physics, Volume 42, pp.2236-2265. ArXiv: math-ph/0005023 (2001).

\bibitem {Quat2} Zhang, F.: Quaternions and matrices of quaternions. Linear algebra and its applications, Volume 251 (1997), pp.21-57. Part of this paper was presented at the AMS-MAA joint meeting, San Antonio, January 1993, under the title \lq\lq Everything about the matrices of quatemions''.

\bibitem {Quat3} Farenick, D., R., Pidkowich, B., A., F.: The spectral theorem in quaternions. Linear algebra and its applications, Volume 371 (2003), pp.75–102.

\bibitem {Hawking} Hartle, J. B., Hawking, S. W.: Wave function of the universe. Physical Review D (Particles and Fields), Volume 28, Issue 12, 15 December 1983, pp.2960-2975.

\bibitem {Bochicchio} Bochicchio, M.: Quasi BPS Wilson loops, localization of loop equation by homology and exact beta function in the large-N limit of SU(N) Yang-Mills theory. ArXiv: 0809.4662 (2008).

\bibitem {minko} Minkowski, H., Die Grundgleichungen fur die elektromagnetischen Vorgange in bewegten Korpern. Nachrichten von der Gesellschaft der Wissenschaften zu Gottingen, Mathematisch-Physikalische Klasse, 53-111 (1908).

\bibitem {einstein3} Einstein, A., Die Grundlage der allgemeinen Relativitatstheorie. Annalen der Physik, 49, 769-822 (1916).

\bibitem {planck} Planck, M., Entropy and Temperature of Radiant Heat. Annalen der Physik, 4, 719-37 (1900).

\bibitem {von} von Neumann, J., Mathematical Foundations of Quantum Mechanics. Princeton University Press, Princeton (1932).

\bibitem {epr} Einstein, A., Podolski, B., Rosen, N., Can quantum-mechanical description of physical reality be considered complete? Phys. Rev., 47, 777-780 (1935).

\bibitem {bridge} Einstein, A., Rosen, N., The Particle Problem in the General Theory of Relativity. Phys. Rev., 48, 73-77 (1935).

\bibitem {bohm} Bohm, D., Quantum Theory. Prentice Hall, New York (1951).

\bibitem {bell} Bell, J.S., On the Einstein-Podolsky-Rosen paradox. Physics, 1, 195-200 (1964).

\bibitem {aspect}Aspect, A., Grangier, P., Roger, G., Experimental realization of Einstein-Podolsky-Rosen-Bohm Gedankenexperiment: a new violation of Bell's inequalities. Phys. Rev. Lett. 49, 2, 91-94 (1982).

\bibitem {spinnet} Penrose, R., Angular Momentum: an approach to combinatorial space-time. Originally appeared in Quantum Theory and Beyond, edited by Ted Bastin, Cambridge University Press, 151-180 (1971).

\bibitem {spinfoam1} LaFave, N.J., A Step Toward Pregeometry I.: Ponzano-Regge Spin Networks and the Origin of Spacetime Structure in Four Dimensions. ArXiv:gr-qc/9310036 (1993).

\bibitem {spinfoam2} Reisenberger, M., Rovelli, C., 'Sum over surfaces' form of loop quantum gravity. Phys. Rev. D, 56, 3490-3508 (1997).

\bibitem {loopgravity} Engle, G., Pereira, R., Rovelli, C., Livine, E., LQG vertex with finite Immirzi parameter. Nucl. Phys. B, 799, 136-149 (2008).

\bibitem {matrix_model} Banks, T., Fischler, W., Shenker, S.H., Susskind, L., M Theory As A Matrix Model: A Conjecture. Phys. Rev. D, 55, 5112-5128 (1997). Available at URL http://arxiv.org/abs/hep-th/9610043 as last accessed on May 19, 2012.

\bibitem {lisi} Garrett Lisi, A., An Exceptionally Simple Theory of Everything. ArXiv:0711.0770 (2007). Available at URL http://arxiv.org/abs/0711.0770 as last accessed on March 29, 2012.

\bibitem {superg} Nastase, H., Introduction to Supergravity ArXiv:1112.3502 (2011). Available at URL http://arxiv.org/abs/1112.3502 as last accessed on March 29, 2012.

\bibitem {Maudlin} Maudlin, T., Quantum Non-locality and Relativity. 2nd ed., Blackwell Publishers, Malden (2002).

\bibitem {spin2} Penrose, R., On the Nature of Quantum Geometry. Originally appeared in Wheeler. J.H., Magic Without Magic, edited by J. Klauder, Freeman, San Francisco, 333-354 (1972).


\end{thebibliography}
\end{document}